# A Multi-Gene Genetic Programming Application for Predicting Students Failure at School

**J.O. Orove**

Department of Computer Science
University of Port Harcourt
Rivers State, Nigeria
E-mail: joshuaorove2012@gmail.com
Tel: +234 7067992097

**N.E. Osegi**

Department of Information and Communication Technology
National Open University of Nigeria
Lagos State, Nigeria
E-mail: geeqwam@gmail.com
Website: www.osegi.com
Tel: +234 7030081615

**B.O. Eke**

Department of Computer Science
University of Port Harcourt
Rivers State, Nigeria
E-mail:bathoyol@gmail.com
Tel: +234 8037049586

**ABSTRACT**
Several efforts to predict student failure rate (SFR) at school accurately still remains a core problem area faced by many in the educational sector. The procedure for forecasting SFR are rigid and most often times require data scaling or conversion into binary form such as is the case of the logistic model which may lead to lose of information and effect size attenuation. Also, the high number of factors, incomplete and unbalanced dataset, and black boxing issues as in Artificial Neural Networks and Fuzzy logic systems exposes the need for more efficient tools. Currently the application of Genetic Programming (GP) holds great promises and has produced tremendous positive results in different sectors. In this regard, this study developed GPSFARPS, a software application to provide a robust solution to the prediction of SFR using an evolutionary algorithm known as multi-gene genetic programming. The approach is validated by feeding a testing data set to the evolved GP models. Result obtained from GPSFARPS simulations show its unique ability to evolve a suitable failure rate expression with a fast convergence at 30 generations from a maximum specified generation of 500. The multi-gene system was also able to minimize the evolved model expression and accurately predict student failure rate using a subset of the original expression.

## 1. INTRODUCTION

SFR has always being and will continue to be a major concern to stakeholders in the educational sector. It refers to the proportion (or more correctly percentage) of students not graduating in a chosen course of study [1]. It is an important aspect of educational curricula assessment as this will help educational administrators to evaluate the performance of their existing curricula, teaching system, infrastructure and student relations programmes. Since the performance of any school system is largely affected by the failure rate of the students, it becomes necessary to study this obviously very important parameter. In particular, there has been a global call to reduce the failure rates of science school students, especially in the Mathematics courses [2]. Social graphs and data mining techniques [3, 4] have been suggested and some cases. Logistic and multiple linear regression techniques have also been used to study student failures rates [5, 14]. Methodologies for investigating student failure rates or decline in academic performance using artificial intelligence techniques such as Neuro-Genetic Algorithms (NGAs), Artificial Neural Networks (ANNs), Genetic Algorithms and decision trees [6,7,8, 9], have been suggested and developed in the literature.

More recently, GP has been applied in a grammar guided genetic programming algorithm to predict if the student will fail or pass a determined course and identifies activities to stimulate its learning in a positive or negative way through the use of True Positive (TP) and True Negative (TN) concepts [10].

### 1.1 Statement of Problem

In our educational institutions today there is currently the need to improve student's performance by reducing the number of students who fail at school. This requirements demand that school administrators and managers employ specific tools, methods and approaches to effectively study and predict student's performance overtime. Currently, such predictive tools are either not flexible, or are very expensive and scarce. This paper seeks to fill this gap by introducing a Multi-gene genetic programming application, GPSFARPS, to facilitate the process of student failure rate analysis and to aid educational administrators predict student's failure at school more cost-effectively.

### 1.2 Theoretical Framework Using the GP Model

The prediction of student failure rate can be carried out using different approaches with consequently different kinds of results but with the sole purpose of evaluating the number of failing students in an exam or course of study. Models such as the Logistic Regression Model [5] have been used to predict the actual probability that a student will pass or fail a Chemistry test. However, when dealing with input data of the continuous form and with a small data set, this approach becomes ineffective leading to effect size attenuation [17]. Also the issue of coding and proper labelling has to be taken into account if the logistic technique is to be utilized as a viable predictor model. Classification and Regression trees [18] have been used to predict students passing or failing a subject taking into consideration six socio-demographic variables (age, gender, ethnicity, education, work status, and disability). However, as the author [18] pointed out in his concluding remarks, limitations do exist with the use of classification trees which could distort results obtained in the classification process and as in the case of the logistic regression, the prevalence of small data sets leads to low prediction accuracies.

More recently, educational data mining techniques [4, 6] have been applied to the growing problem of predicting student failure rate with high degree of success rates.

### 1.3 Research purpose

We intend to bridge the gap in this area, which is to develop a software application that can evolve a model expression for predicting student failure at school given a history of class/exam scores- called the historical data set. These model expressions can then be easily utilized by the school administrators in carrying out their predictions or failure rate calculations for a future data set.

### 2. RELATED LITERATURE

Prediction is one of the most frequently studied problems on Data Mining and Machine Learning researchers. It consists on predicting the value of a categorical attribute based on the values of other attributes, the predicting attributes. One of the most useful Data Mining tasks on educational data mining is classification [16]. It makes possible the prediction/classification of a student's performance, for example. There are different educational objectives for using classification, such as: discovery of potential student groups with similar characteristics and reactions to a particular pedagogical strategy[19], to detect student's "misuse or game-playing" the system which is correlated with substantially lower learning [27], to group students that are hint-driven or failure driven and find common misconceptions that students possess [20], to identify learners with a low rate of motivation and find remedial actions to lower drop-out rates [21], to classify/predict students when Intelligent Tutoring Systems (ITS) are being used [22]. To predict student outcome, some studies have been made: prediction of student's grades, on a scale from A to F, from test scores using neural networks (counter propagation networks and back propagation networks) [23]; prediction of the relevance of classes, i.e., if classes are relevant or not to student's academic success, using discriminant function analysis [24] Predicting a student's academic success (low, medium or high risk) using decision trees, random forests and neural networks [25].

Using GeSES [26], a method that has been designed specifically to work with students logs, based on C4.5 rules, a main goal was established by its authors: to test if symptoms of a bad adaptation in an adaptive course where detected. A bad adaptation was detected if the course generated a low performance, caused by bad student's inadequate environment. Genetic programming is also a broadly used method on Educational Data Mining. G3PMI [10], a grammar guided genetic programming algorithm, has been applied to predict if the student will fail or pass a determined course and identifies activities to stimulate its learning in a positive or negative way through the use of True Positive (TP) and True Negative (TN) concepts. Results show that G3P-MI achieves better performance with more accurate models than a better trade-off between such contradictory metrics as sensitivity and specificity. Also, it adds comprehensibility to the knowledge discovered and finds interesting relationships between certain tasks and the time expended to solving exercises with the final marks obtained in the course.

Tanner [28] used a k-nearest neighbour (kNN) method to predict student performance in an online course environment. Extensive experimental results from a 12-lesson course on touch-typing, with a database of close to 15000 students were presented. The results indicate that kNN can predict student performance accurately, and already after the very first lessons. They also concluded that early tests on skills can be strong predictors for final scores also in other skill-based courses. Kalles [9] used Genetic Algorithms and Decision trees for a posteriori analysis of tutoring practices based on Student's failure models. Their results showed that genetic induction of decision trees could indeed produce very short and accurate trees that could be used for explaining failures. In [10], data mining techniques and real data of about 670 middle-school students from Zacatecas, México were used to predict school failure. Several experiments were carried out in an attempt to improve accuracy in the prediction of final student performance and, specifically, students that were likely to fail.

In the first experiment the best 15 attributes was selected. Then two different approaches (Data Balancing and Cost-Sensitive approaches) were applied in order to resolve the problem of classifying unbalanced data by rebalancing data and using cost sensitive classification. The outcomes of each one of the approaches using the 10 classification algorithms and 10 fold cross validation were compared in order to select the best approach to the problem. From the results, it was deduced that OneR fared better with a TN (True Negative – Fail) rate of about 88.3% when using data balancing approach, whereas Jrip fared better with a TN rate of 93.3% with the cost sensitive approach. They examined the differential effects of prior academic achievement, psychosocial, behavioural, demographic, and school context factors on early high school grade point average (GPA) using a prospective study of 4,660 middle-school students from 24 schools and a combined Multiple Linear Regression (MLR)/ Hierarchical Linear Modeling (HLM) approach. Their findings suggest that:
(a) Prior grades and standardized achievement are the strongest predictors of high school GPA and
(b) Psychosocial and behavioural factors (e.g., motivation, self-regulation, and social control) add incremental validity to the prediction of GPA. When comparing the relative importance of each set of predictors (the dominance analysis technique), the variance accounted for by psychosocial and behavioural factors is comparable to that accounted for by prior grades.

These findings highlight the importance of effective risk assessment based on multiple measures (i.e., academic, psychosocial, and behavioural) for the purpose of identifying risk, referring students to intervention, and improving academic success [10].

They also used a new form of genetic programming called Grammar based Genetic Programming (G3P), combined with an Interpretable Rule Classification Mining Scheme (ICRM) for the prediction of student failure rate at school using real data from high school students in Mexico. These data were based on class marks and of high dimensions (variable intensive) and imbalanced, thus, the need for variable (dimension) reduction, cost sensitive classification/re-sampling of original datasets respectively to reduce these irregularities was emphasized. They compared three versions of their G3P/ICRM model with 10 classification algorithms to evaluate the Fail (or True Negative) rates and they found out that the best results were those obtained by their G3P/ICRM models.

However, one thing that is missing in most of the techniques is having a structured user friendly application that will facilitate the generation of a symbolic model expression for describing student failure rate over a particular period of time. Such expressions can make the job of the educational administrator easier particularly in the area of re-computation using a supportive regression approach.

## 3. SOFTWARE METHODOLOGY

In this section, a multi-gene genetic programming approach is employed. This approach uses an iterative radical unified modelling process wherein In particular we have used an open source framework for genetic programming GPTIPS® [11], as a basis for the developed GP application. The system is robust, modular and customizable with user friendly interfaces.

### 3.1 Theoretical Foundations

GP operates based on the Darwinian principle of evolution, natural selection, and survival of the fittest. Typically, GP solutions are evolved individual programs encoded in a structure referred to as a gene or gene tree. At the beginning of a GP algorithm the genes or expression trees are randomly initialized within a feasible solution space, and then they undergo reproduction, crossover and mutation.

Reproduction is a technique used to replicate new genes from the original parent genes akin to sexual reproduction in the programming tournament. The programming tournament can be likened to be an environment of competing programs where the best of them is selected as the eventual winners.

The two core genetic operations during reproduction are [13]:
  (i) Crossover: This involves the interchange of genetic material among the solutions or genes
  (ii) Mutation: This involves random changes (additions and deletions) within a gene itself.

The crossover operation employs a swapping function in software while the mutation operation uses a knock-out technique.

### 3.2 Genetic Programming Process

The fundamental mathematical formulations guiding a genetic programming application and corresponding pseudo-codes are not new and are given in [12]. In general a GP system seeks to minimize the mean square error of the fitted data set by evolving multiple solutions at different intervals of time. A typical architecture of a GP Programming sequence is shown in Fig1.

A genetic programming process starts from providing genetic programming with basic building blocks of the solution and some method of analyzing how well a proposed solution solves the problem. This is also followed by supplying the Genetic programming with the fitness metrics which the Genetic programming (GP) will use in generating solution. In Fig 1 the process analysis of the genetic programming is presented and it shows clearly that the first set of solutions generated does not become the final solution. Instead the solutions keep evolving over and over certain number of times till an optimal solution is arrived at. In the figure it is clear that the GP begins with some initial guess at a solution and successively attempts to improve the solution over time.

Once some criterion for termination (either an ideal individual or some predefined run time) GP returns the best individual so far. That individual is deemed the *result of GP or the Optimal Solution.* In Fig 2 the more detail diagram shows what happens internally inside the Genetic Programming box. It begins by generating some initial population. The fitness of all individuals in the population is then assessed. It is unlikely that this initial generation will contain an ideal individual, but some will likely be better than others.

We begin the GP loop by selecting the individuals that solve the problem best and allowing them to reproduce, making small random changes to their construction. As this process repeats numerous times, we find that on average, the fitness of the population tends to increase.

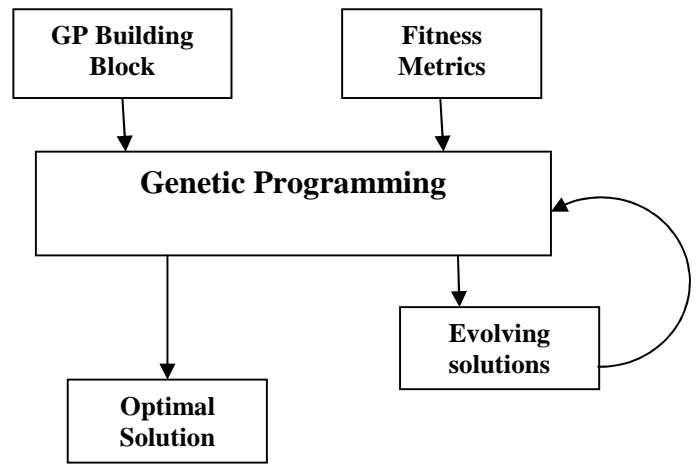

Fig1: Simplified Genetic Programming Process Analysis

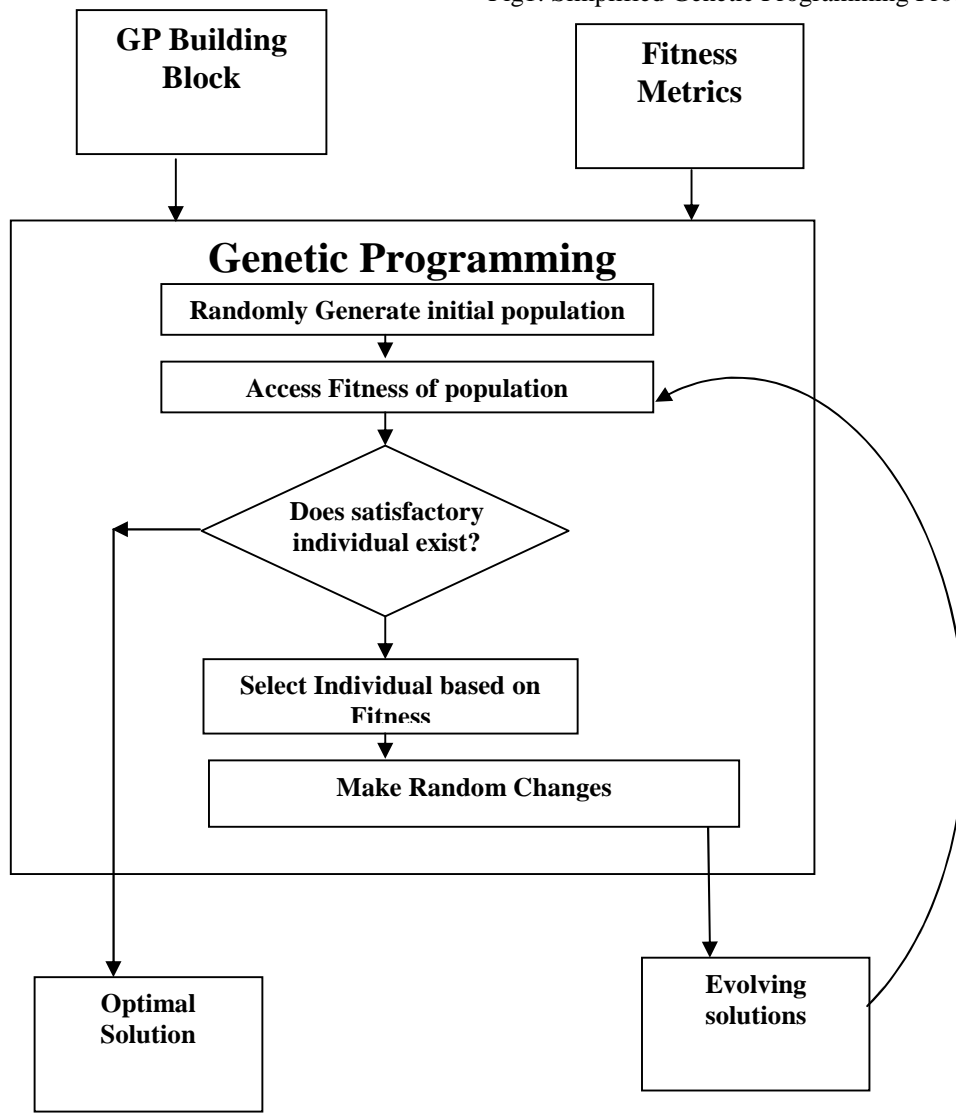

Fig2: Main Genetic Programming Process Analysis

### 3.3 GP Model Solutions and Methodology

The first step in the design of a Genetic Programming is the design of the problem representation. In this section we will present the representation design of the system.

### 3.3.1 Representation (Building Block)

In our analysis above we have been talking a lot about individuals, for instance "Select Individual based on Fitness" but not really discussing what these individuals are made of. We know that GP aims to evolve computer programs, but what kind of computer programs? The computer programs GP evolves are programs written in some functional programming language. In functional programs, the idea is to take some input and simply return a value without dealing with computer state. You can think of functional programming more like mathematical expressions than instructions for somebody to follow. There are many functional programming languages, including LISP and OCaml. Another option is to represent the individuals using objects of whatever programming language that are been used to code the GP system. This is the way our Darwin GP Environment handles representation. However whether we choose to represent individuals in some functional programming language or with objects in memory, GP evolves individuals that can be represented as a tree structure. It is more functionally useful to consider them in this way when thinking about GP. The tree structure design in Fig 3 is the building block for our genetic programming.

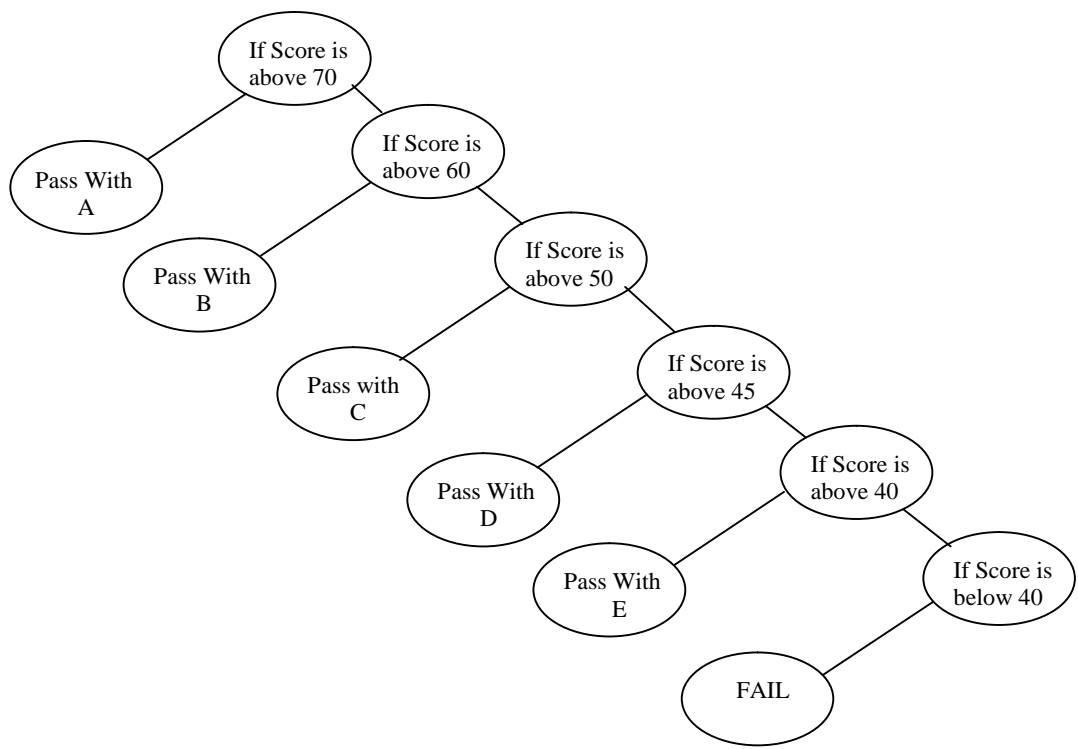

Fig3: A Tree Representation of the Student Failure GP

### 3.3.2 Multiple Gene Symbolic Regression Genetic Programming (GP) Model:

A multigene individual consists of one or more genes, each of which is a "traditional" GP tree. Genes acquired incrementally by individuals in order to improve fitness (e.g. to reduce a model's sum of squared errors on a data set). The overall model is a weighted linear combination of each gene. Optimal weights for the genes are automatically obtained (using ordinary least squares to regress the genes against the output data). The resulting pseudo-linear model can capture non-linear behaviour.

In multigene symbolic regression based GP, each prediction of the output variable y is formed by the weighted output of each of the trees/genes in the multigene individual plus a bias term. Each tree is a function of zero or more of the N input variables x1, … xn.

Mathematically, a multigene regression model can be written as:

$$y = d_o + d_1*tree1+...+d_m*tree_M \quad\quad (1)$$

where $d_o$ = bias (offset) term and $d_1,..., d_m$ are gene weights and M is the number of genes (i.e. trees) comprising the current individual. The weights (i.e. regression coefficients) are automatically determined by a least squares procedure for each multigene individual.

The number and structure of the trees is evolved automatically during a run (subject to user defined constraints) using training data, i.e. a set of existing input values and corresponding output values. Testing data (another set of input and corresponding output values from the process or system you are modelling) can be used, after the run, to evaluate the evolved models. The testing data is not used to evolve the models and serves to give an indication of how well the models generalise to new data.

A pseudolinear multigene model of predictor output y with inputs $x_1$ to $x_6$; the weights $d_0$, $d_1$, $d_2$ are automatically obtained by least squares.

**Multiple gene model (transfer function):**
$$Y = d_o + d_1*x1 + d_2*x2 + d_3*x3 + d_4*x4 + d_5*x5 + d_6*x6 \quad \ldots\ldots\ldots (2)$$

The architecture of the developed system is given in Fig 4.

### 3.3.3 Solution Representation and Methodology

Genetic programming creates computer programs in the Lisp or scheme computer languages as the solution. Genetic algorithms create a string of numbers that represent the solution. In our model, an inductive-evolutionary approach is employed. The developed application based on Genetic programming uses the basic executional steps as follows to solve the failure rate problem:

- Step 1: Input data file containing RxC matrix of input (predictor) variables, an Rx1 matrix of output (response) variables, and system configuration file is fed to the GP system
- Step 2: Generate a random initial population of expressions based on Step1 using a multi-gene model
- Step 3: Assign a fitness value to each individual in the population
- Step 4: Create a new population of model expressions
- Step 5: Choose the best existing solutions
- Step 6: Create new solutions by crossover and mutation
- Step 7: The best solution is chosen as the fittest solution (least error margin).
- Step 8: Based on the best solution expression apply grammar based classification to establish the number of students below failure value
- Step9: Compute failure rate and print solution or candidate failure rate expression

Fig 4 shows the architecture of a typical Multi-Gene GP sequence. It must be emphasized here that Step1 specifically builds the structure of the GP system. Step 2 to 7 is achieved using the rungp functional class [11].

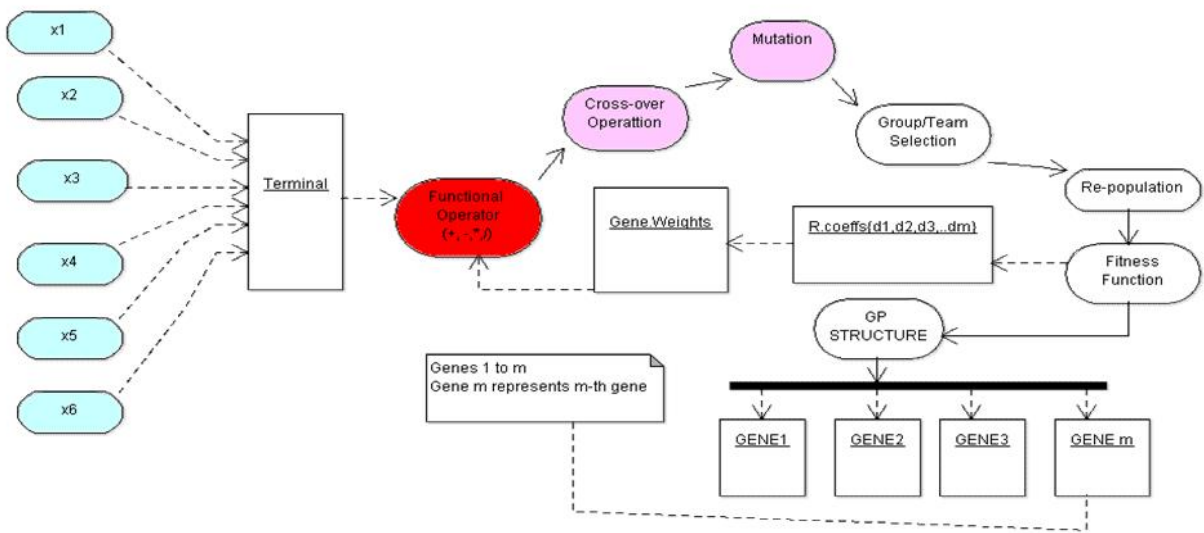

Fig 4: Architecture of a typical Multi-Gene GP Operation

### 3.3.4 Application Development

The application has been built in phases. The development phases use the concepts of windowing and navigation common to most modern GUI's. Windows are basically clickable user interfaces that facilitate data entry, retrieval and visualization in a compact and easy to use way using components such as buttons, text fields, menus, etc. The application is designed and developed in phases using callback functions or classes within a scripting program. It basically consists of three core windows (not shown):

**Start window:** serves as front end for application

**Main window:** for specifying core GP parameters and running the GP prediction system

**Grammar Window:** for failure rate classification and estimation. A display has been added to the Main window and Grammar window to print the solution model expression for prediction and failure classifications respectively.

It is possible for the user to move between windows by using callback buttons or menus. Fig 5 shows the class diagram including the core computational functions used in the developed application.

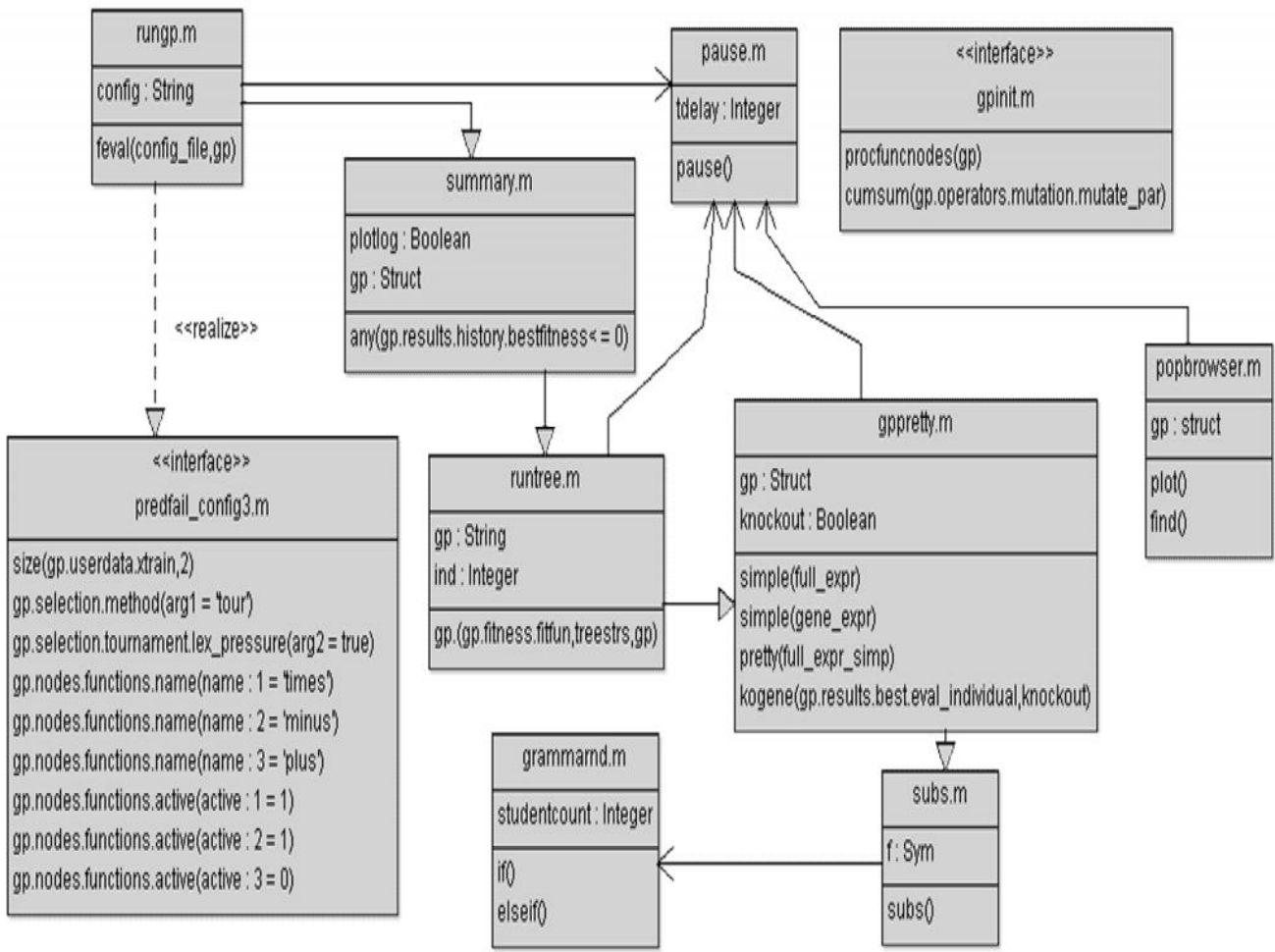

Fig 5: Computational Class Diagram of GPSFARPS

### 3.3.2 GP Model Input Data and Specifications

A continuous set of student raw scores used in the developed GP applications are proposed with sample student net scores labelled Q1 to Q5, and net continuous assessment, labelled CA score, jointly referred to as the Predictor Variables, while the output parameter (response variable) is the total score (TOTAL) which is as usual is graded and scaled to a maximum of 100 and minimum of 0. These data were obtained from historical records and are provided in Table 1. We have made the assumption that the scores represent a time series so future forecasts can be easily achieved and assured even when there is great variation in size, the principles remain the same i.e. we could easily scale down to a suitable range. In Table2, the GP parameters for the analysis and prediction process are presented taking into account bloat constraints [8].

Table1:
Sample of Student Raw Score Data

| Time | x1 | x2 | x3 | x4 | x5 | EXAM SCORE | C.A. SCORE (x6) | TOTAL | GRADE |
|------|----|----|----|----|----|------------|-----------------|-------|-------|
| 1 | 0 | 7 | 0 | 0 | 0 | 7 | 16 | 23 | F |
| 2 | 6 | 7 | 0 | 6 | 5 | 24 | 20 | 44 | E |
| 3 | 1 | 6 | 3 | 0 | 0 | 10 | 16 | 26 | F |
| 4 | 0 | 4 | 4 | 3 | 2 | 13 | 17 | 30 | F |
| 5 | 0 | 4 | 2 | 2 | 1 | 9 | 16 | 25 | F |
| 6 | 0 | 5 | 0 | 9 | 6 | 20 | 20 | 40 | E |
| 7 | 0 | 2 | 1 | 1 | 0 | 4 | 16 | 20 | F |
| 8 | 0 | 4 | 0 | 0 | 0 | 4 | 16 | 20 | F |
| 9 | 5 | 10 | 0 | 0 | 5 | 20 | 20 | 40 | E |
| 10 | 0 | 11 | 4 | 6 | 5 | 26 | 21 | 47 | D |
| 11 | 8 | 4 | 2 | 0 | 0 | 14 | 16 | 30 | F |
| 12 | 0 | 8 | 6 | 6 | 0 | 20 | 21 | 41 | E |
| 13 | 0 | 7 | 0 | 0 | 4 | 11 | 16 | 27 | F |
| 14 | 0 | 2 | 0 | 0 | 0 | 2 | 16 | 18 | F |

Table2: GP Parameter Settings

| GP algorithm parameters | Parameter Settings |
|-------------------------|--------------------|
| Population size | 150 |
| Number of generations | 500 |
| Selection method | Plain Lexicographic Tournament Selection |
| Tournament size | 4 |
| Termination criteria | 0 |
| Maximum depth of each tree | 4 |
| Maximum number of trees | 4 |
| Maximum Number of Genes | 4 |
| Mathematical operators for symbolic regression | $\{+, -, x\}$ |

## 4. RESULTS

In this study GPSFARPS application is used to train the raw score data from table 1 with specifications given in Table 2. A lexicographic tournament selection strategy is chosen for selecting the parent genes from the pool of available solutions. The tournament size is set to 4. The maximum depth of each tree in the multi-gene representation is set to 4 and this allows some control over the complexity of the evolved expressions. The instruction set or the functions that are used for symbolic regression are $\{+, -, x\}$.

Figure 6 shows the final population of the GP run showing the trade-off between the accuracy of the fit and the complexity of the evolved models. The solutions points labelled in blue represent the set of dominated solutions while those in green represent the set of non-dominated solutions or the Pareto front. Solutions which are on the Pareto front are non-dominated in the sense that there are no solutions which have both a lower fitness and a lower fitness and a lower model complexity simultaneously than these ones. In other words, if another solution has a lower fitness value then it must have a higher model complexity and vice-versa [12]. Fig 7 also captures the individuals involved in the best solution space.

The MGGP convergence characteristics in Figure 8 also indicate that, 30 generations is sufficient for the convergence of the algorithm. It can be seen that the objective function does not change significantly near the end of the GP run. Also the associated curve of the mean fitness is plotted below it. It shows that the overall population loses diversity very quickly and running the GP algorithm for more number of generations is not going to yield a much better solution. However, it can also be seen that the algorithm should not be run for less than 10 generations for the present case, as the solution would not have converged sufficiently by then.

The generated modelled expression is captured in Fig 9
which can be written as
Y = x6 + x5 +x3 + x4 ……………. (3)

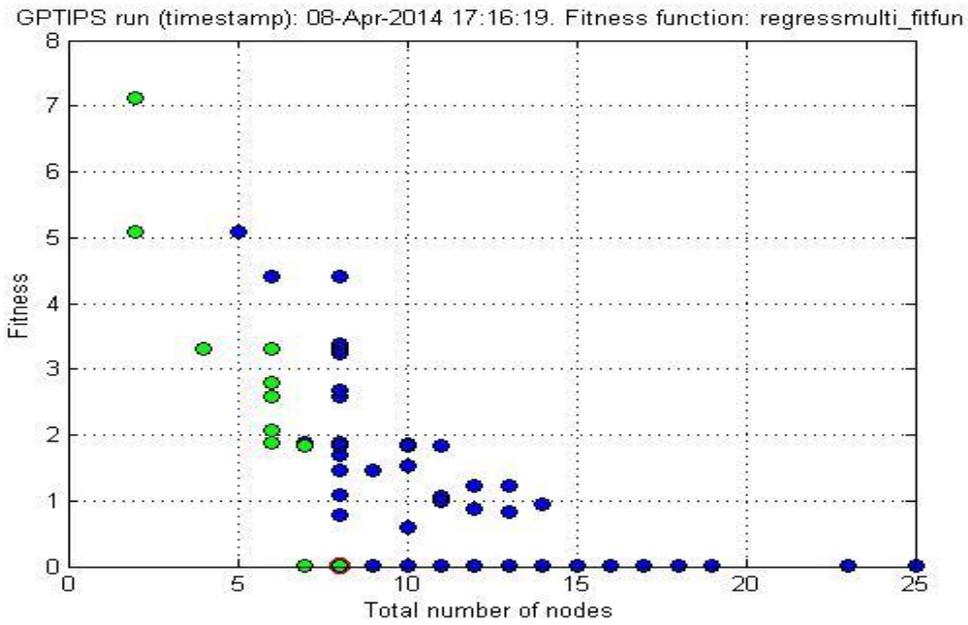

Fig 6: Fitness vs. complexity of the evolved Multi-Gene
GP solutions along with Pareto solutions.

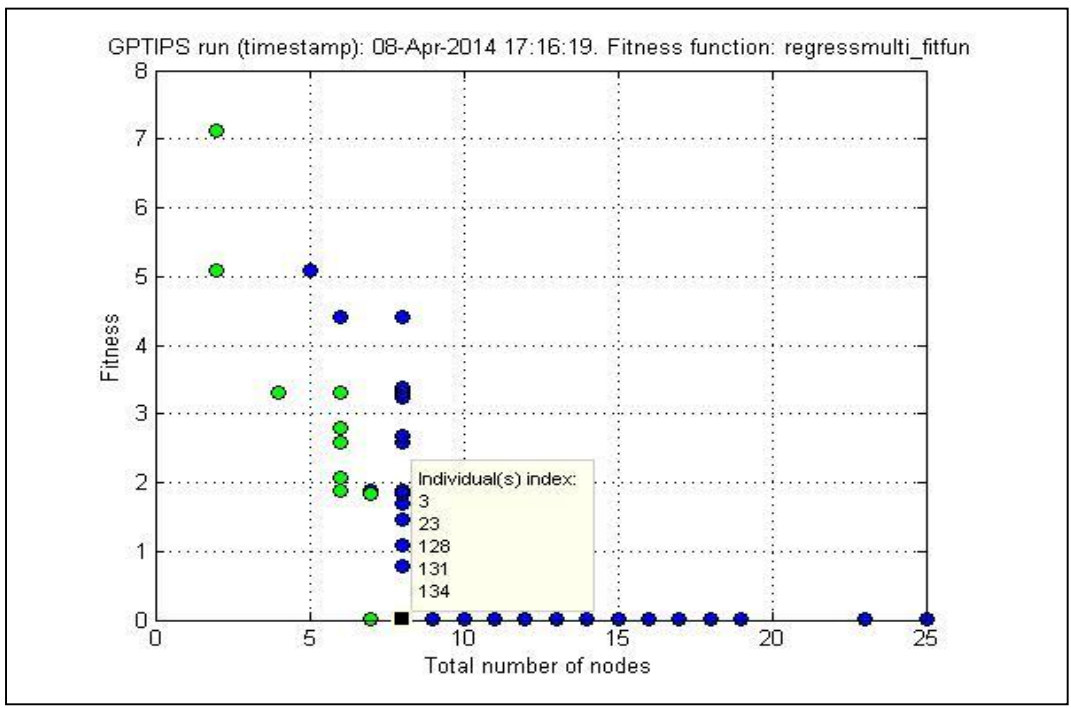

Fig 7: Fitness vs. complexity of the evolved Multi-Gene GP solutions along with
Pareto solutions and best individuals (3, 23, 128, 131, 134).

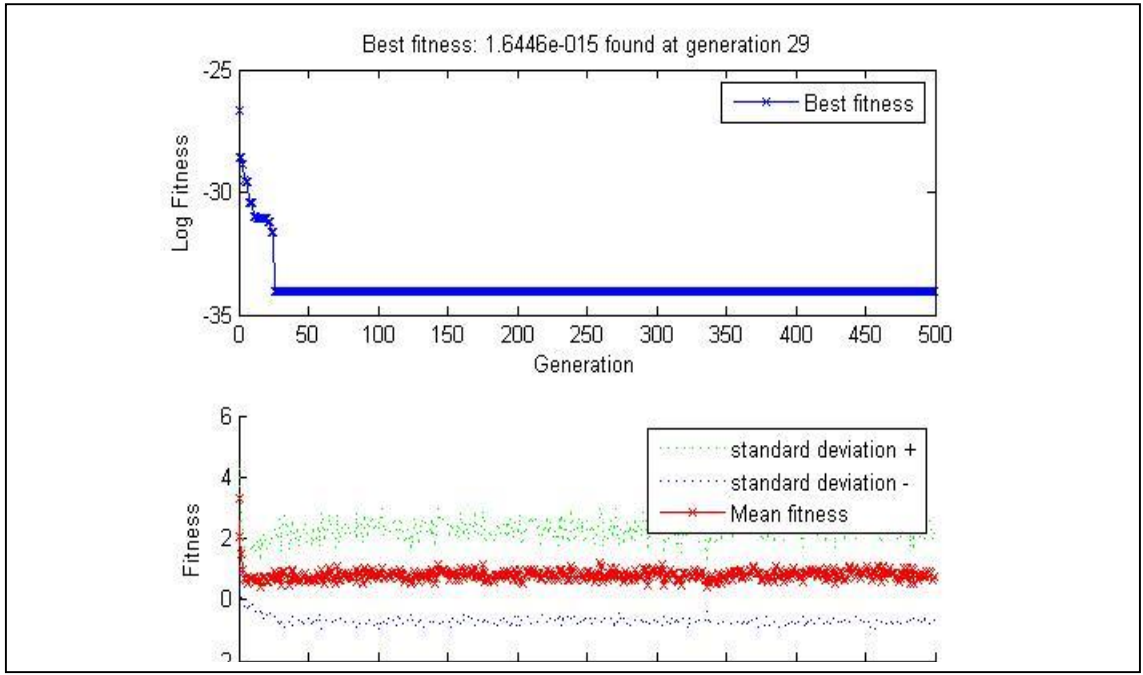

Fig 8: Convergence curves for the entire Multi-gene GP run with the best and mean fitness of solutions

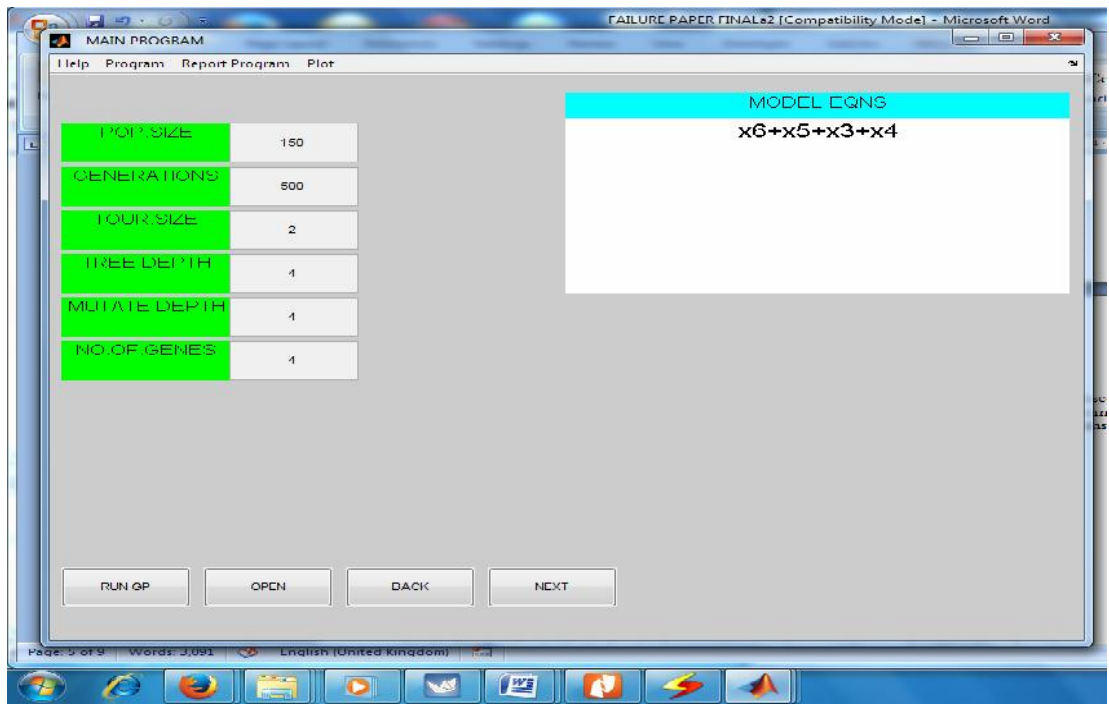

Fig 9: Snapshot of Generated model equation in GPSFARPS

## 5. DISCUSSIONS

In this study we have implemented the principles of GP using the multi-gene approach in a friendly user interface application. We applied the concept of cross-over and mutation to evolve a random population of individual solution programs and obtained a best fitted model for the specified GP parameters. The graphical reports are generated using MATLAB © handle graphics technology [15]. Using the interactive software GPSFARPS, GP parameters can be specified and meaningful features observed at run time.

A parsimonious model can be achieved if the core GP parameters, in particular, the number of generations, population size, tournament size, tree depth, number of gene trees and number of genes are constrained. From Fig 8, we have shown that specifying arbitrary number of generations does not significantly improve the system performance beyond possible achievable limits (Log Fitness versus Generation Plot). We should also point out here that the problem of local minima common to other AI schemes such as neural networks and fuzzy logic can be avoided with this evolutionary technique. Fast convergence has been achieved at approximately 30 generations from a maximum of 500 generations. This fast convergence rate may be attributed to the close correlation between predictor and response variables but should not be expected at all times. Also four candidate response variables, $x_6$, $x_5$, $x_4$, and $x_3$, were generated by the system and then used in the prediction process. This clearly shows the strength of multi-gene GP in minimizing or simplifying complex predictor equations.

Another issue what considering is the need for flexible specification of functional operators or function data set(s), since the function set plays a vital role in the structure of the evolved expression, such a technique can be valuable if the school administrator has first hand knowledge of the shape of the failure rate. The failure shape or failure rate graph may be discovered using simple graph operations on test data from which an initial guess for suitable candidate functions may be arrived at. Once candidate function(s) have been deduced, it can then be programmed into GPSFARPS using end-user programming call back functions. In this way, novel solutions to the failure rate problem may be discovered by the end-users themselves.

## 6. CONCLUSIONS

A object-oriented evolutionary software application for evolving model expressions that can predict student failure rates at school has been developed. The system was fed with incomplete raw score data and simulated response was satisfactory with a fast convergence rate – about 30 generations for a specified maximum generation of 500 and a minimal model expression consisting of only 4 predictor variables from a total of six predictor variables. The system can safely be used to facilitate predictions on complex data by using the software interface. The application will be open sourced and made available on the web to facilitate further development.

## 7. SIGNIFICANCE AND RECOMMENDATIONS FOR FUTURE WORK

This paper has presented a software application for predicting student failure rate at school. The software is capable of regenerating failure rate model expressions using a time –series dependent student's raw scores as input parameters and these model expressions can in turn be used by school administrators to determine student failure rate at school

The software can be made more useful if a bloat checking mechanism is incorporated. Further investigations need to be carried to test the effect of two state logic data on the GP software system developed, improving the function set database and making the program platform independent. Also, One-step in this direction will be to build a suitable software framework that will encourage end-user programming of the function data set. An interactive-agile function data set to facilitate end user programming of GPSFARPS in a future design is thus proposed.

## REFERENCES

[1] Annelize J, 2009. "The Failure Rate of Hospitality Management Students in Win Courses at Tshwane University of Technology: Exploring Probable Causes. Faculty of Management Sciences Tshwane University of Technology, South-Africa

[2] Sarah J.P, Overcoming Poor Failure Rates in Mathematics for Engineering Students: A Support Perspective. Harper Adams University College, Harper Adams Learning and Teaching Support Services, pp. 1-8

[3] Fire M, Katz G., Elovici Y, Shapira B, and Rokach L, 2012. Predicting Student Exam's Scores by Analyzing Social Network Data. AMT'12 Proceedings of the 8th international conference on Active Media Technology Springer-Verlag Berlin, Heidelberg, pp. 584-595, doi:10.1007/978-3-642-35236-2_59

[4] Vera C.M., Romero C., and Ventura S., 2011. Predicting Student Failure Using Data Mining, Journal of Applied Intelligence, , MA, USA, vol. 38 (8), Kluwer Academic Publishers Hingham, MA, USA, pp. 315-330.

[5] Johnson, W.L., Johnson, A.M., and Johnson J., 2012 Predicting Student Success on the Texas Chemistry STAAR Test: A Logistic Regression Analysis, http://eric.ed.gov/?id=ED534647.

[5] Johnson, William L., Johnson, Annabel M., and Johnson, Jared. Predicting Student Success on the Texas Chemistry STAAR Test: A Logistic Regression Analysis, http://eric.ed.gov/?id=ED534647, 2012

[6] Randall S. S, 2001. Neural Networks Refined: using a Genetic Algorithm to Identify Predictors of IS Student Success, The Journal of Computer Information Systems, vol. 41, Issue 3, pp. 42-47.

[7] Huang S., 2011. Predictive Modeling and Analysis of Student Academic Performance in an Engineering

Dynamics Course. All Graduate Theses and Dissertations.Paper1086, http://digitalcommons.usu.edu/etd/1086.

[8] Huang S., and Fang, N., 2013. Predicting student academic performance in an engineering dynamics course: A comparison of four types of predictive mathematical models, Journal of Computers and Education, Elsevier, vol. 61, pp. 133-145.

[9] Kalles D., Pierrakeas C., 2006. Using Genetic Algorithms and Decision Trees for a posteriori Analysis and Evaluation of Tutoring Practices based on Student Failure Models, in IFIP International Federation for Information Processing, vol. 204, Artificial Intelligence Applications and Innovations, Boston: Springer, pp. 9-18.

[10] Rquez-Vera C.M., Cano A., Romero C., and Ventura S., 2013. Predicting student failure at school using Genetic Programming and different data mining approaches with high dimensional and imbalanced data, Springer Science Business Media, vol. 38, pp. 315- 330, doi: 10.1007/s10489-012-0374-8.

[11] Searson P., 2009. GPTIPS: Genetic Programming & SymbolicRegressionforMATLAB, http://gptips.sourceforge.net.

[12] Indranil P, Daya S.P and Saptarshi D., 2013. Global solar irradiation prediction using a multi-gene genetic programming approach, Journal of Renewable and Sustainable Energy, vol.5, no.6, pp.1-30, doi: 10.1063/1.4850495.

[13] Riccardo P., William B.L., Nicholas F.M., and Koza J.R, 2008. A Field Guide to Genetic Programming, pp. 2

[14] Jesuína M., Fonseca B., and Joseph E. C., 2006. Secondary Student Perceptions of Factors affecting Failure in Science in Portugal, Eurasia Journal of Mathematics, Science and Technology Education, pp. 82-95

[15] Matlab version R2007b, 2007. Mathwork s Inc., USA, http://mathworks.com

[16] Romero C., Ventura S., Espejo P.G., and Hervás C., 2008. Data Mining Algorithms to Classify Students, Proceedings of the 1st International Conference on Educational Data Mining, Montréal, Québec, Canada, June 20-21, 2008

[17] Elite Research, 2013. Multiple Regression, Kean University, pp. 1-8

[18] Zlatko J.K., 2011. Predicting student success by mining enrolment data, Research in Higher Education Journal, pp.1-20.

[19] Chen, G. 2000. Discovering Decision Knowledge from Web Log Portfolio for Managing Classroom Processes by Applying Decision Tree and Data Cube Technology, Journal of Educational Computing Research

[20] Yudelson, M.V., 2006. Mining Student Learning Data to Develop High Level Pedagogic Strategy in a Medical ITS, AAAI Workshop on Educational Data mining, pp. 1-8

[21] Cocea, M., and Weibelzahl, S., 2006. Can Log Files Analysis Estimate Learners' Level of Motivation?, Workshop on Adaptivity and User Modeling in Interactive Systems, Hidelsheim, pp. 32-35

[22] Hamalainen W., Vinni, M., 2011. Classifiers for educational data mining., Chapman & Hall/VCRC London.pp 120- 340

[23] Fausett, L., Elwasif, W., 1994. Predicting Performance from Test Scores using Backpropagation and Counter propagation, IEEE Congress on Computational. pp. 112 - 320

[24] Martinez, D., 2001. Predicting Student Outcome using Discriminant Function Analysis, Annual Meeting of the Research and Planning Group, California, pp 163-173

[25] Superby, J.F., Vandamme, J.P., Meskens, N., 2006. Determination of Factors Influencing the Achievement of the First-year University Students using Data Mining Methods, Workshop on Educational Data Mining, pp. 37-44.

[26] Bravo, J., 2009. Checking the Reliability of GeSES: Method for Detecting Symptoms of Low Performance, Proceedings of the 9th International Conference on Intelligent Systems Design and Applications.

[27] Baker, J., 2002. Developing a generalizable detector of when student game the system, User Modeling and User-Adapted Interaction, April 02

[28] Tanner T., and Toivonen H., 2010. Predicting and preventing student failure – using the *k*-nearest neighbour method to predict student performance in an online course environment, Dept of Comp.Sc Helnsiki, Finland, pp.1-23